\documentclass{article}
\usepackage{hq2kpro}
\begin{document}
\title{REVIEW OF HEAVY-QUARK PRODUCTION AT FIXED-TARGET EXPERIMENTS
}
\author{Jeffrey A. Appel                               \\
{\em Fermilab, P.O. Box 500, Batavia, IL 60510, USA}   \\
}
\maketitle
\baselineskip=11.6pt
\begin{abstract}
An increasingly large amount of quality fixed-target data on heavy-quark
production at fixed-target energies is appearing.  This data can
provide information across a range of physics topics.  The topics vary
from investigations of QCD predictions to the understanding of the
structure of hadrons.   Recent results on neutrino, photon, and
hadron production of charm and beauty will be reviewed in this
context.  The greatest insight will come from combining multiple
measurements as they relate to the physics topics, and by ensuring that
the parameters used in models are consistent with all the measurements.  
We have not yet really entered the time when this has been done for
fixed-target measurements.
\end{abstract}
\baselineskip=14pt
\section{Introduction}
Heavy-quark production at fixed-target energies provides information
and lessons across a range of physics topics.  These topics are not
so much related to the nature of heavy quarks as such, but rather to
our understanding of basic QCD theory, both perturbative and 
non-perturbative, and to measurements of the nature of hadrons.  The
results of fixed-target experiments are also necessary to interpret the
results of heavy-ion experiments as they search for evidence of
quark-gluon plasma.  The measurements made by fixed-target experiments
can be listed simply enough:\\

\indent Cross Sections: Total, and as Functions of Beam Energy and
Species,\\
\indent \indent $x_F$, and $p_t$\\
\indent Dependence of Production on Nuclear Target A Value\\
\indent Final State Ratios, Particle/Antiparticle Ratios\\
\indent Polarization\\
\indent Correlations Among Heavy-Quark Particles in the Final State\\

\noindent Each of these measurements relates to one or more of the
physics topics.   However, the greatest understanding will come from
combining multiple measurements,
and by ensuring that the parameters used in models are consistent with
all the measurements - not just a single measurement at a time.  We have
not yet really entered the time when this has been done for fixed-target
measurements.  The closest we come is when perturbative QCD calculations
and Monte Carlo simulations with default parameters are compared to
data.  However, the default parameters have yet to be tuned across the
bulk of modern measurements.
\section{Today's Relevant Experiments and Results for This Meeting}
The most relevant fixed-target experiments for the physics topics of
interest are listed in Table \ref{expts}.  These experiments cover a
variety of beam particle types at a range of energies, and use a
plethora of target materials.  Fortunately, many production measurements
are little affected by the details of target nucleus, and we have
learned how to relate measurements on different nuclear targets, at
least for inclusive cross sections.  Measurements such as production
asymmetries, which are self normalizing, have yet to show target
dependences, for example.  And, the total charm cross-section
measurements on nuclei appear to scale with the number of nucleons in
the target nucleus.\cite{E789charmA,E769charmA,WA82charmA}   However, in
specific kinematic regions, and for the small fraction of charm
production which is ``onium,'' the A-dependence is more
complicated.\cite{JAdep,YAdep}  See the discussion below for this latter
effect.
\begin{table}
\centering
\caption{\it Current Fixed-Target Heavy-Quark Experiments of Relevance}
\vskip 0.1 in
\begin{tabular}{|c|c|c|c|} 
\hline
Experiment       &Beam       & Beam     & Target Material \\
                 & Momentum  & Particle &                 \\
                 & (GeV/$c$) &          &                 \\
\hline
\hline
E690             & 800       & $p$      & $LH_2$                    \\ 
\hline
E771             & 800       & $p$      & Si                        \\
\hline
E866/NuSea,      & 800       & $p$      & $LH_2, LD_2$, C, Ca, Fe, W\\
E789 and E772    &           &          & Ag, Au, and Cu dump       \\
\hline
E769       & 250 &$\pi^{\pm}, K^{\pm}$, and $p$ & Be, Al, Cu, and W \\
\hline
E781/SELEX       & 600       & $\Sigma^-$ and $\pi^-$ & C and Cu    \\  
                 & 572       & $p$                    & C and Cu    \\
\hline
E791             & 500       & $\pi^-$                & C and Pt    \\
\hline
E815/NuTeV       & 20 to 400 &$\nu_{\mu}$, ${\overline \nu}_{\mu}$&Fe\\
\hline
E687             & 220       & $\gamma$               & Be          \\ 
\hline
E831/FOCUS       & 170       & $\gamma$               & BeO and Si  \\
\hline
WA89             & 340       & $\Sigma^-$ and $\pi^-$ & C and Cu    \\
\hline
WA82             & 340       & $\pi^-$                & Si, Cu, and W \\
                 & 370       & $p$                    & Si and W    \\
\hline
WA92/Beatrice    & 350       & $\pi^-$                & Cu and W    \\
\hline
\end{tabular}
\label{expts}
\end{table}

Several measurements are newly available for this conference, and others
have only just been published or submitted for publication.  These recent 
and new results are listed by experiment in Table \ref{results}.
The various results just beginning to appear from FOCUS, SELEX, and
NuTev from the the 1996-7 Fermilab fixed-target run bode well for
continuing, and even more interesting results in the near future.  
Nevertheless, it is a good time to review what we have learned so far
from the heavy-quark fixed-target experiments of the past.

\begin{table}
\centering
\caption{\it New Heavy-Quark Production Results for This Conference}
\vskip 0.1 in
\begin{tabular}{|c|l|} \hline
Experiment       & Measurement \\
\hline
\hline
E690        & Diffractive Production of $D^*$                         \\
E771        & $c$ and $b$ Production by Protons                       \\
E791        & $D^*$ Production: $x_F, p_t^2$, Polarization, Asymmetry \\
E791        & $\Lambda_c$ Production Polarization and Asymmetry       \\
E781/SELEX  & Production Asymmetries                                  \\
E831/FOCUS  & Charm Particle Correlations                             \\
E815/NuTeV  & Neutrino Production of Charm                            \\
E866        & $J/\psi, \psi'$, and $\Upsilon$ Production and A-Dependence\\
\hline
\end{tabular}
\label{results}
\end{table}

\section{Heavy-Quark Production Mechanisms and Measurements}
Heavy-quark production is interesting for two reasons.  First, the
lifetime and uniqueness of the flavor of heavy quarks allows us to
follow the progress of a single quark from its production to its
emergence as a fully developed hadron observable in the laboratory.
Thus, we can probe the time development of hadronic processes involving 
heavy quarks.  Secondly, since the production of heavy quarks by photons
and hadrons is so dominated by the gluon content of projectiles (via
photon-gluon fusion in photoproduction and gluon-gluon fusion in
hadroproduction), the study of heavy-quark production allows the
investigation of the gluon content of incident hadrons, both the mesons
and baryons in beams and the nucleons in targets.  In the case of
neutrino interactions, the production of charm is dominated by
W-exchange off strange quarks in the nuclear sea.  Thus, we can also
learn about strange sea quark distributions in nucleons.  In addition,
charm production in neutrino Deep Inelastic Scattering is an important
ingredient in tests of two-scale perturbative QCD, the two scales being
the charm mass and $\Lambda_{QCD}$.


We have implicitly assumed that the production process can be divided
into separate considerations of the incident partons, their inelastic
interaction producing heavy quarks, and the hadronization process of
these quarks.  This division is referred to as factorization.  In the
following discussion, each observation will be relevant to one part of
the factorized process.


In heavy quark production at fixed-target energies, a single process
tends to dominate.  For charm quark production, these processes are
neutrino-strange-quark charged current scattering, photon-gluon fusion,
and gluon-gluon fusion, respectively for incident neutrinos, photons,
and hadrons.  In each case, the target parton density plays a direct
role, and measurements of each process is sensitive to that parton
density.  Thus, the strange quark sea distribution can be measured in
neutrino charm production, and gluon densities in photo- and
hadroproduction of charm.


In neutrino production of charm, E815/NuTev is just now starting to show
results on sea-quark distributions.  They have found that the strange
and non-strange sea quark distributions have a small asymmetry at most,
and that the strange sea is about 40\% of the non-strange sea.
These results are discussed in more detail in the talk of Maxim
Goncharov at this Workshop.

\begin{figure}[b]
\vspace{6.5cm}
\includegraphics{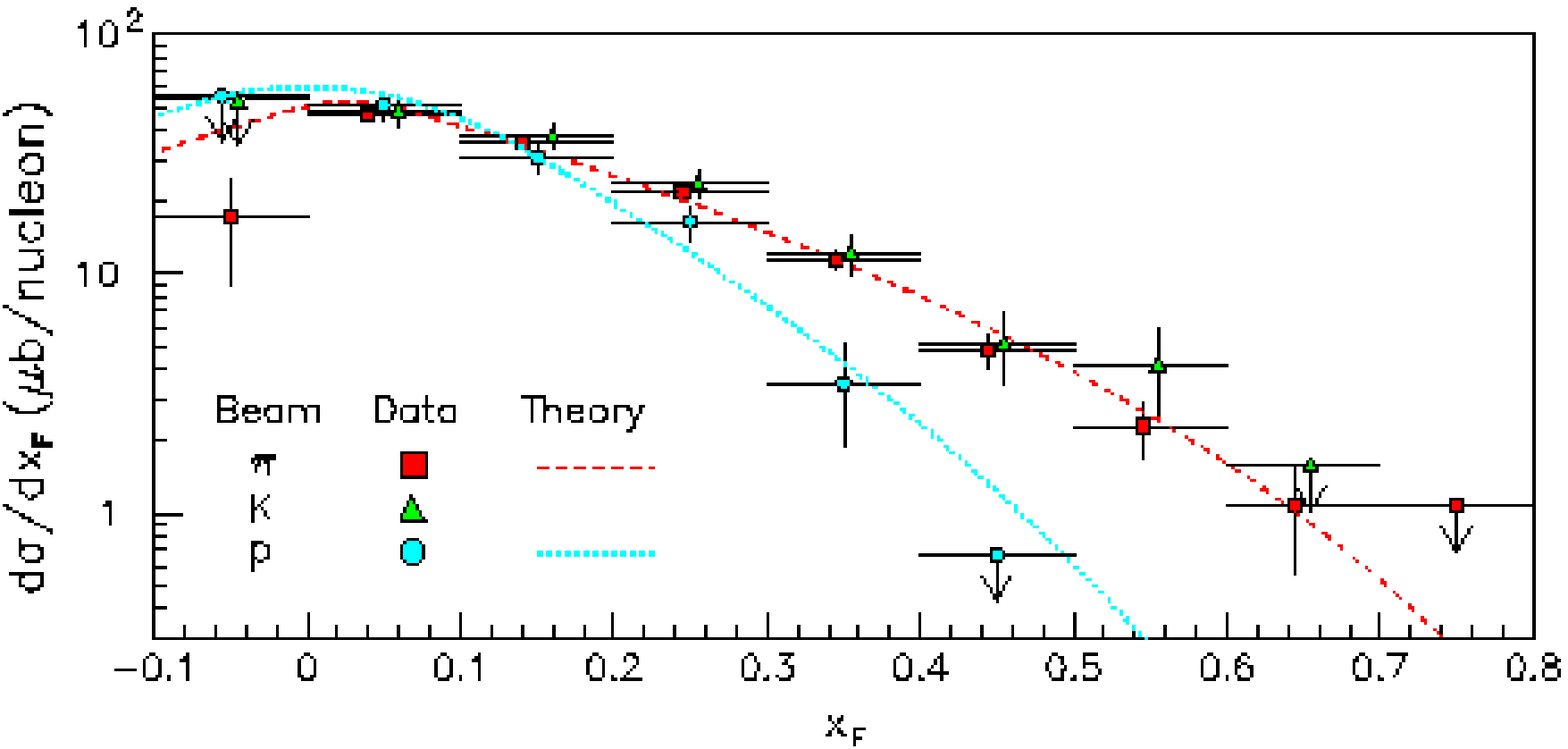}
\caption{\it
      E769 inclusive charm particle $x_F$ distributions showing $\pi$
and $K$ meson beams leading to similar results, different from those for 
incident protons.
    \label{e769xF} }
\end{figure}

From E769, we have rather direct evidence that the gluon densities in
mesons are harder (i.e., have higher momentum fractions on average) than
those in nucleons.\cite{E769xF}  This harder gluon momentum leads to
more forward production of D mesons when the incident particle is a
meson than when it is a proton, as shown in Fig. \ref{e769xF}.  Note
that the pion and kaon appear to produce the same charm particle $x_F$
distribution, implying that the gluons in pions and kaons are the same,
i.e., SU(3) symmetric.  This is the same sort of kinematic argument
which explains why the photoproduced charm particles are even more
forward.  The photon interacts with its entire momentum, pushing the
subprocess center of mass more forward than typical partons in hadrons,
where only a fraction of the hadron momentum goes to each parton.

In the intrinsic charm model of Stan Brodsky, Romona Vogt, and
collaborators, a virtual charm quark-antiquark pair appears among the
sea quarks of a hadron, and is knocked onto the mass shell during an
interaction.\cite{brodsky}  The diagram for this process is explicitly
included in higher order perturbation theory.  What is different in this
model is that the process is pulled out explicitly for calculation, with
the initial charm sea quarks given by a parton distribution function.  
The intrinsic charm was thought to account for 1 to 2$\%$ of the proton,
an amount required to explain early CERN Intersecting Storage Ring
experiment results.  These intrinsic charm pairs tend to be co-moving
with the valence quarks of their parents.  So, it is easy for such
quarks to coalesce with valence quarks.  This coalescence causes the
same asymmetry as would be expected for any other coalescence
process; i.e., it should appear at high $x_F$.  However, we would also
expect it to be limited to low $p_t$.  The evidence is that production
asymmetries are rather flat in $p_t$;\cite{E769asy,E791asy,E791tuneMC}
so no direct evidence for intrinsic charm exists here.  In addition, the
differential cross section for $J/\psi$ production also limits the size
of intrinsic charm, to less than 1\% {\it of the prediction} for their
Be target.\cite{E789Jpsi}.  An earlier measurement in 800 GeV/$c$
proton-Si interactions by E653\cite{E653} were interpreted as giving
a $0.2\%$ upper limit on the amount of intrinsic charm in the proton.

\section{Cross Sections}


The total charm cross section (typically taken as dominated by the
inclusive ground-state meson production cross section) has been measured
in photoproduction and hadroproduction over energies from about ten
GeV to hundreds of GeV.  Cross sections have also been calculated in
leading order (LO) and next-to-leading order (NLO) in perturbative
QCD.  The ratio of NLO to LO cross sections is sometimes referred to as
the ``K-factor.''  This factor is significantly greater than one,
typically a factor of a few.  Nevertheless, it is considered likely that
next-to-next-to-leading order terms will not be so important.  The NLO
differential cross sections, like the total cross sections, appear to be
related to the LO calculations by more-or-less a single number, the same
K-factor.  So, shapes of distributions do not change much, and the
absolute values are usually played down compared to relative
differential cross sections (shapes).  

The range of fixed-target photoproduction results are extended to 40 TeV
equivalent photon beam energy by measurements at the HERA collider.  
Even these highest-energy photoproduction data fit a NLO calculation
over the full range quite well.\cite{Berezhnoy}  Note, however to watch
the inclusion of color-octet effects at low photon energy and the
parameters used in the cited calculation.  Even so, the theoretical
uncertainties in photoproduction are rather smaller than those for
hadroproduction, where uncertainties associated with the charm-quark
mass and the appropriate scale for the calculations each give
uncertainties of nearly an order of magnitude.  The pattern of
cross-section measurements agree with each other across the full range
of experiments and energies quite well.\cite{E791prod}


The total beauty cross section at fixed-target experiments has been
measured at various energies for both incident pions and protons.
The measurements usually come from incomplete B decay observations;  
e.g., $J/\psi$'s which do not come from the primary interaction, and
are presumed to be from $B$ decays.\cite{E771mumu}  The data exists for
incident pions and protons, and is quite consistent with the exception
of one measurement with incident 800 GeV/$c$ protons.  Unfortunately,
even for the heavy bottom quark, the NLO QCD calculations have an
uncertainty of a factor of nearly ten.

It is worth noting explicitly that all the heavy-quark-production
calculations use a universal running value of $\alpha_S$.  
Any discrepancy between comparisons of calculations of charm and
bottom data could have been evidence for a breakdown of QCD.  However,
in this context it is worth noting the limits on such non-universality
of the strong coupling are strongly limited by measurements from
$e^+e^-$ collisions.\cite{SLDalpha}

\begin{figure}[t]
\vspace{6.5cm}
\includegraphics{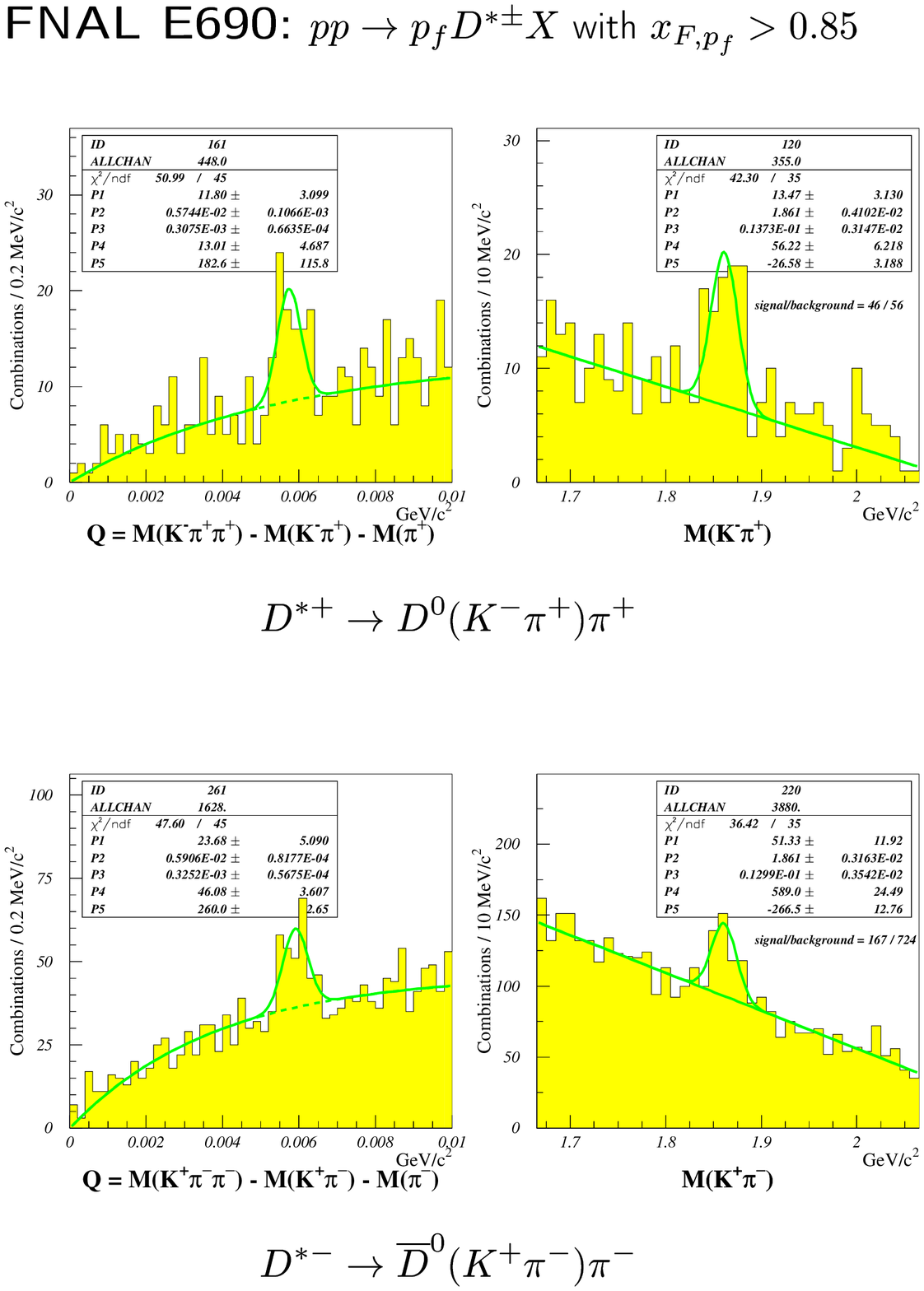}
\caption{\it
      E690 diffractive $D^*$ signals, using the usual decay to $D^o$
$\pi$.  The data come from $p$ $p$ $\rightarrow p_{forward}$ $D^*$ $X$
with $x_{F,p_{forward}} > 0.85$.
The $D^{*+}$ signal is on top, $D^{*-}$ on the bottom.  The
decay Q-value is shown on the left for the $D^o$ mass-peak regions on
the right. 
    \label{e690diffraction} }
\end{figure}

Another production topic of interest is the size of diffractive charm
production.  In the old days, several CERN and Fermilab experiments
tried to see charm in diffractive events.  It was hoped that this
process would be quite large, perhaps 10\% of the total charm
photoproduction, for example.  Now, we have a first measurement of the
process, but for incident protons, from E690.  The experiment triggers
on a fast, forward proton and looks for $D^*$ mesons (See Fig. 
\ref{e690diffraction}.).  Their measurement of the total diffractive
$D^*$ production cross section is model dependent, ranging from 0.17 to
0.29 $m$b ($\pm$ 0.05 $mb$ statistical error).  The total charm
diffraction is, thus, about 0.75 $mb$, only about 2~$\%$ of the total
charm cross section.  No wonder the early experiments requiring large
diffractive charm production saw so few charm decays.

\begin{figure}[t]
\vspace{6.5cm}
\includegraphics{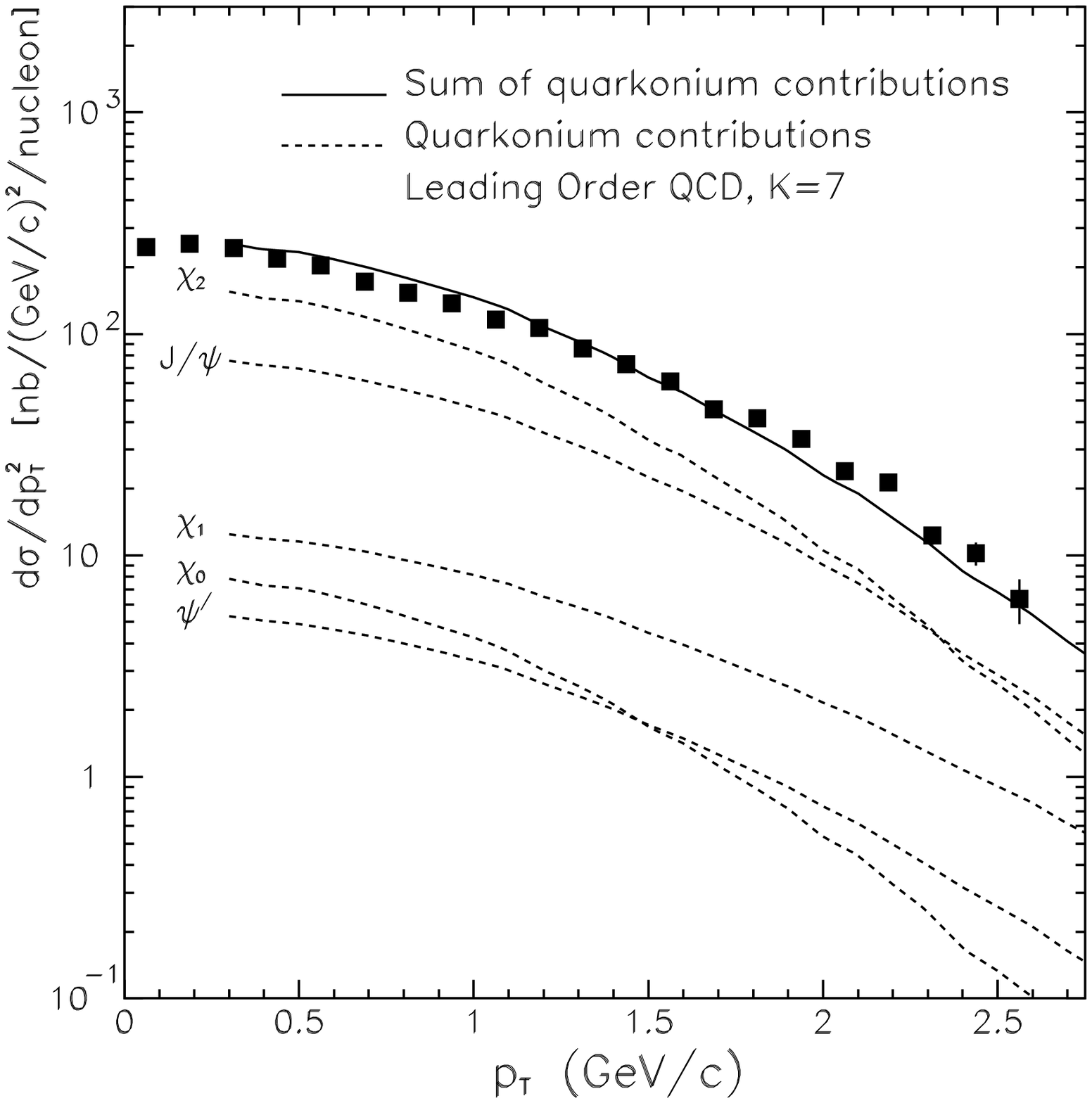}
\includegraphics{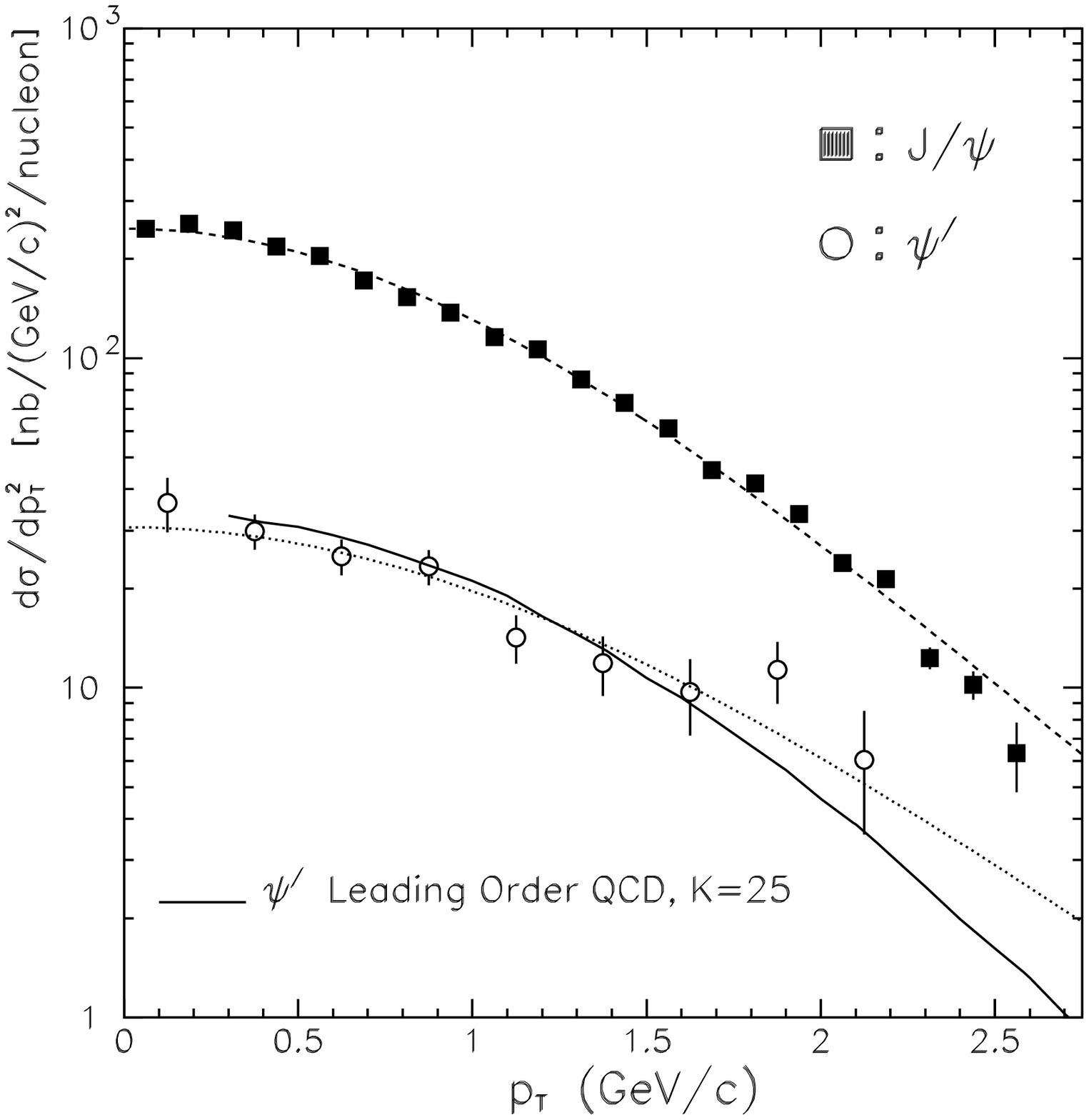}
\caption{\it 
   E789 data on $J/\psi$ and $\psi$' production cross sections and
NLO calculations times large K factors chosen to match the data.
    \label{e789prod} }
\end{figure}

The charmonium cross sections are a real mystery, both at fixed-target
energies and at the Tevatron Collider.  There is data on $J/\psi$, 
$\psi$', $\Upsilon$, $\chi_1$, and $\chi_2$.  The well-publicized,
but unexplained large Collider direct $J/\psi$ and $\psi$' production
(6x and 25x greater than LO calculations, respectively - really large
K-factors!) occurs also at fixed-target energies, and is shown from E789
in Fig. \ref{e789prod} where the production is 7x and 25x larger,
respectively, than the LO calculations!\cite{E789Jpsisigma}  Can the
fixed-target $\chi_1$ and $\chi_2$ production shed light on this?

There are three production models considered in the literature.  Each
makes a prediction on the relative $\chi_1$ and $\chi_2$ production in
$p$ $N$ interactions.  The three models, and their predictions are:\\

   Color singlet $-$ $\chi_1 \sim 5 \%$ of $\chi_2$ production.

   Color evaporation/color bleaching $-$ $\chi_1$:$\chi_2$ production
3:5.

   Non-Relativistic QCD (color octet plus color singlet) $-$ $\chi_1$ \\ 
\indent \indent up to $30 \%$ of $\chi_2$ production.\\

\noindent The inclusion of color-octet processes seems to explain the
observations in hadroproduction.\cite{E771Jpsisigma}  Yet, the
color-octet matrix elements from the Tevatron don't work at HERA.  Further
complicating the situation is that the expected polarization of
color-octet produced charmonium seems not to be present at the Tevatron 
Collider!\cite{CDFpol}  


Some of the most beautiful data of recent years comes from the series of
experiments in Fermilab's Meson East beam line.  Very high statistics
measurements of dimuon production give impressive signals for Drell-Yan
pairs and the heavy-quark onia states as 
well.\cite{JAdep,E789Jpsi,E866Adep}
Among their results is a study of the A-dependence of open charm
\cite{E789charmA} (which agrees with the $A^1$ results of others) and of
$J/\psi$ and $\psi$' (which are of higher detail and precision).  

\begin{figure}[t]
\vspace{6.5cm}
\includegraphics{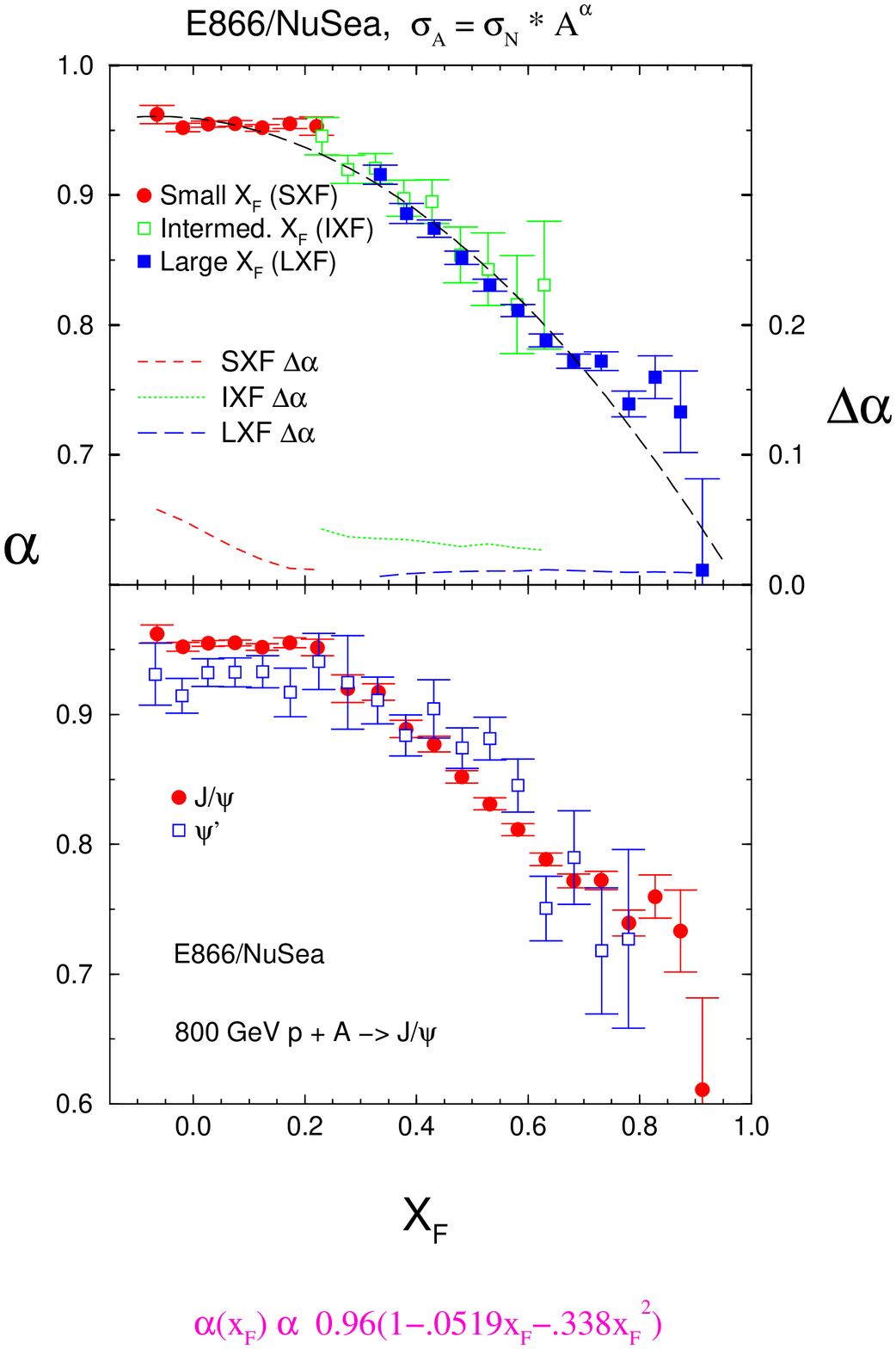}
\includegraphics{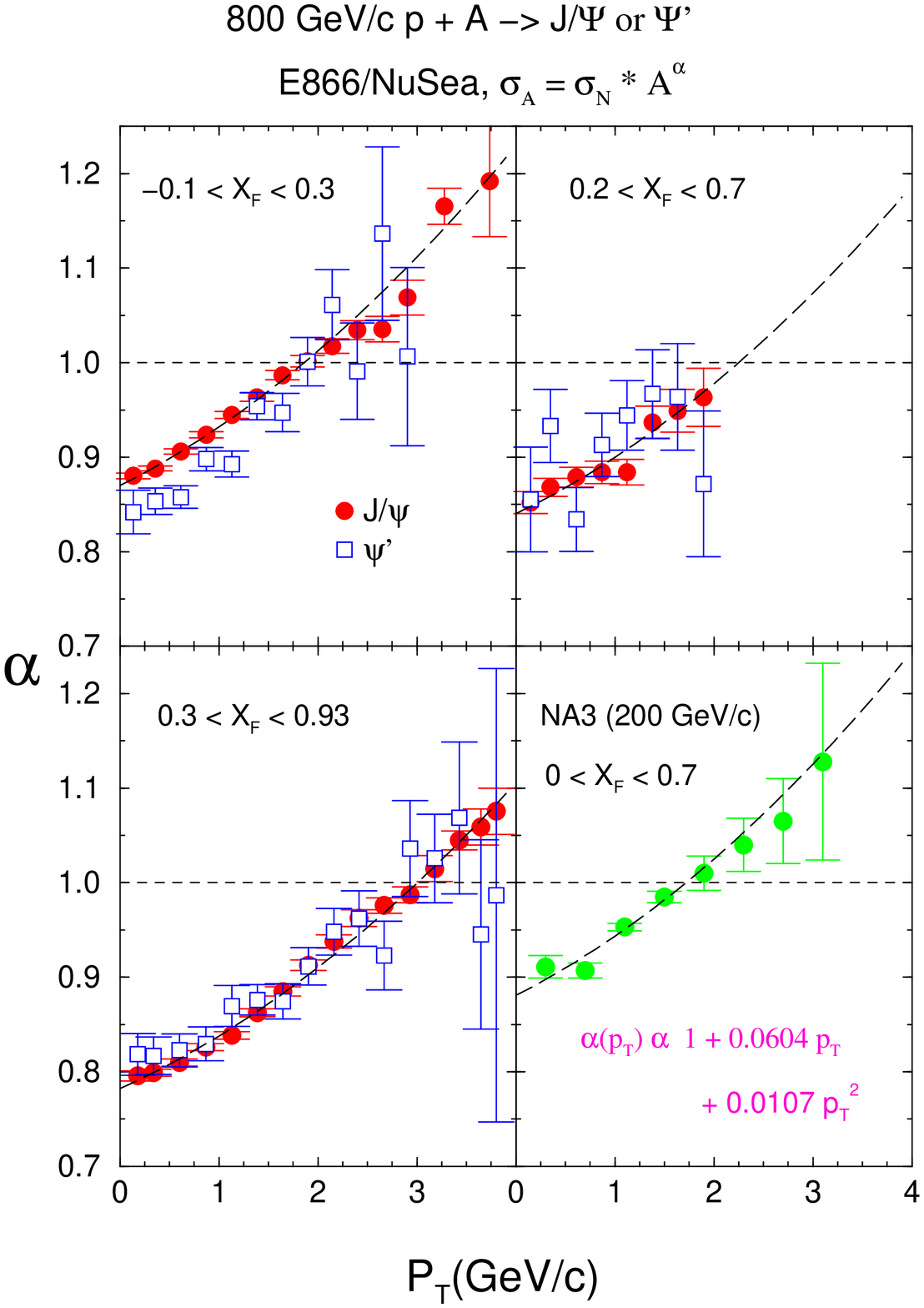}
\caption{\it
       E866 data on $J/\psi$ and $\psi$' production A-dependence showing
rather nonuniform values of $\alpha$, the power of A in the cross
section. 
    \label{e866Adep} }
\end{figure}

As shown in Fig. \ref{e866Adep}, the A-dependence for $J/\psi$ and
$\psi$' production is not simple, varying by kinematic region, though
quite similar for $J/\psi$ and $\psi$'.  Are the details of the $x_F$
and $p_t$ dependences evidence for color-octet production?  Another
feature of these data is the cautionary note sounded against too easy
interpretation of $J/\psi$ production effects in heavy-ion collisions as
evidence for quark-gluon plasma.  The data shown in Fig. \ref{e866Adep}
demonstrate that using a single power $\alpha$ in predictions cannot be
a complete model.  In fact, who is to say that the dense nuclear matter
does not mix the kinematic regions we think of from fixed-target
measurements.  What values of $\alpha$ are relevant?

In these discussions of cross sections, we have an example of the need 
to subject model calculations to the full range of observations before
anointing any model.  So far, this has only been done piecemeal, and
no model seems to explain all the observations.
\section{Hadronization/Fragmentation}
In the factorization scheme, the final stage involves turning the
produced heavy quarks into the hadrons seen in the laboratory.  This 
process is usually referred to variously as fragmentation and
hadronization.  The term fragmentation carries with it an implication of
independence of the heavy quark and antiquark in the process, and the
use of ``fragmentation functions'' measured in $e^+e^-$ collisions.
However, the environments of hadronic collisions and of $e^+e^-$
collisions are quite different in terms of the totality of color fields
present.  We will see that this difference is quite evident, and the
term hadronization appropriately conveys a more complex situation.

\begin{figure}
\vspace{9.5cm}
\includegraphics{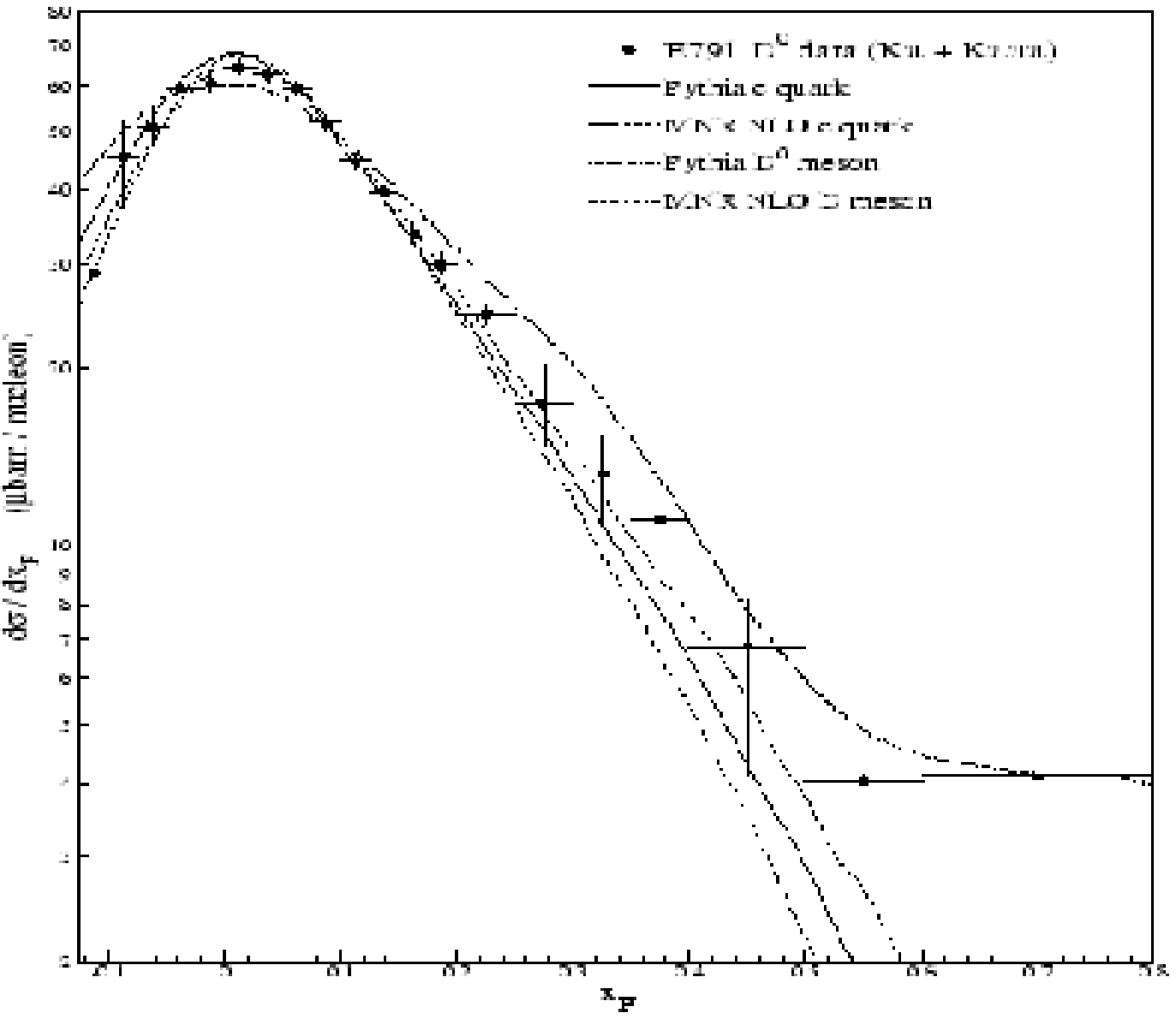}
\caption{\it E791 inclusive $D^o$ production $x_F$ distribution.  
The meson data match the predictions for quarks better than that for
mesons.
    \label{e791xF} }
\end{figure}

Longitudinal momentum distributions of heavy-quark particles are usually
presented as functions of the scaled variable, Feynman x, $x_F$.  This
scaling is relative to the maximum kinematically allowed value, meaning
that the variable runs from $- 1.0$ in the backward direction to $+ 1.0$
in the forward direction.  Plotting distributions versus $x_F$ makes the
distributions measured at various energies look about the same.  More
interestingly, and as shown in Fig. \ref{e791xF}, the observed $D$ meson
$x_F$ distributions look like the QCD predictions for quarks.\cite{E791prod} 
The $x_F$ distribution does not look like the prediction for mesons when
fragmentation functions are used to go from quarks to hadrons.

Why should observed hadrons appear to have more momentum than you might
expect after fragmentation?  There must be some process other than
simple fragmentation which contributes forward momentum to balance what
would be lost by fragmentation.  Such a process has been called ``color
drag.'' It refers to the color string attachment between the produced
heavy quarks and the remnant quarks of projectiles.  The forward-moving
remnants ``pull'' the heavy quark forward during the hadronization
process.  As we will see below, production asymmetries lend credence to
this idea.


In hadroproduction, the longitudinal distributions of heavy-quark
particles and antiparticles may differ even though the leading order
QCD process is symmetric in quark and antiquark production.  In fact,
even including next-to-leading order processes only produces a very
small asymmetry in the far forward direction.  What is observed in
experiments, for example, is large asymmetries when a heavy-quark
particle has a valence light-quark in common with the incident hadron
and the antiparticle does not.\cite{E769asy,E791asy,E791tuneMC,E781asy} 
This is known as the ``leading particle effect.''  Of course, it should
also refer to asymmetries in the backward direction when heavy-quark
particles have light valence-quarks in common with the target hadrons.

The effect is understood as being due to the coalescence of the heavy
quark with a valence quark from projectile (or target) when the heavy
quark and light quark are close in phase space.  The experimental
evidence supports this picture across a wide variety of charm particles
and incident beam particle types.
        
There are also particle/antiparticle asymmetries in photoproduction.
In the photon direction, or centrally, the process must have another
explanation.  There, we may be seeing the effects of associated
production (meson with baryon).  The energy threshold for such
associated production is less than for heavy baryon-antibaryon
production.  And, we may expect that this associated production effect
is also responsible for some of the asymmetry seen in hadroproduction,
especially in the central region (near $x_F$ of zero).
 
The coalescence/recombination model seems to provide a framework for
understanding particle/antiparticle production asymmetries.  Intrinsic
charm is not apparently required.  Leading-particle effects increase in
the forward and backward directions, according to the expectations tied
to valence-quark content.  

A particularly interesting feature of the asymmetries is that they are
more or less flat in $p_t$.\cite{E769asy,E791asy,E791tuneMC,E781asy}
This feature appears in the PYTHIA/JETSET simulations, but is not
expected in the intrinsic quark model as suggested above.  It should be
noted, however, that the default parameters of current versions of the
PYTHIA/JETSET software do not get the details of the $x_F$ and $p_t$
asymmetry dependences right (see below).


Experimenters typically report particular measurements in their papers,
rather than presenting more universal coverage of results, even their
own.  When making comparisons, they may vary the input parameters of
models to find those that provide the best match to the particular
measurement being presented.  E791, for example, in comparing their
$D^{\pm}$ production asymmetry to PYTHIA predictions, show both the
default parameter predictions and those with modified
parameters.\cite{E791tuneMC}  In particular, the charm quark mass and
intrinsic $k_t$ (see discussion of this parameter below) are changed
from 1.35 GeV/$c^2$ and 0.44 GeV/$c$ to 1.7 GeV/$c^2$ and 1.0
GeV/$c$.  The paper notes that these parameters are not unique in
obtaining agreement, and that it is necessary to select a set of
parameters which fit this and a range of other measurements to have any
confidence in the parameters.  

\begin{figure}[t] 
\vspace{4.5cm}  
\includegraphics{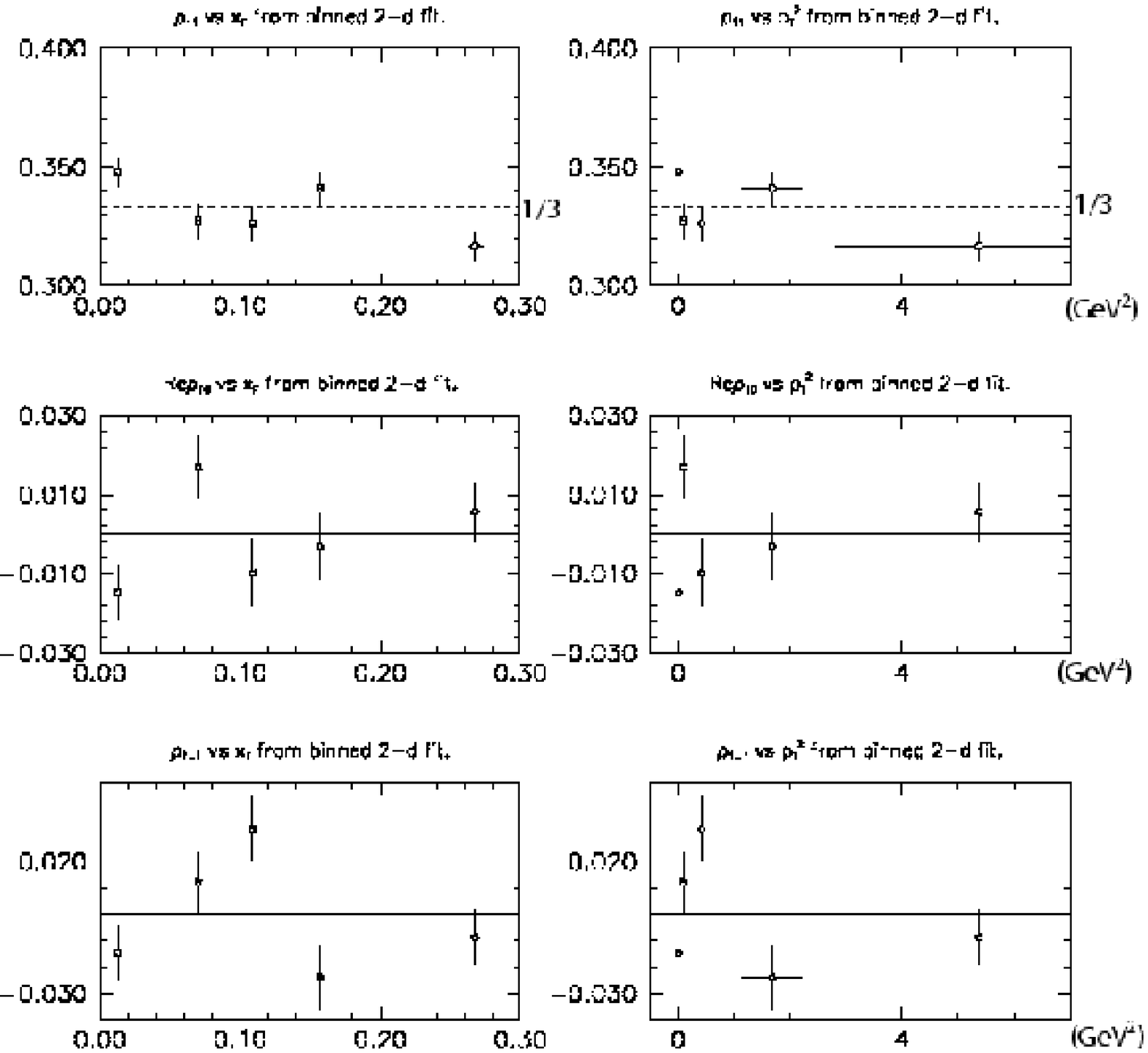}
\includegraphics{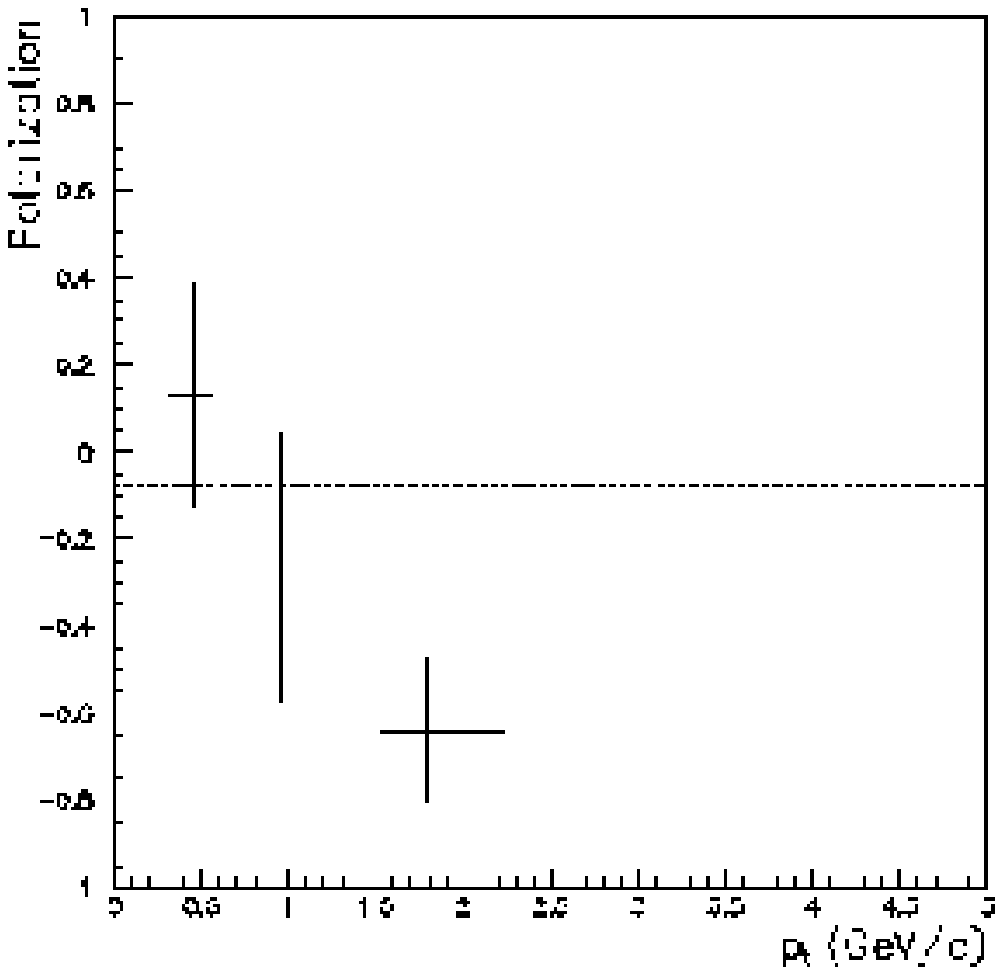}
\caption{\it  
      E791 $D^*$ spin-density matrix elements vs $x_F$ and $p_t$ (left)
and $\Lambda_c$ polarization vs $p_t$ (right).
    \label{e791Pol} }
\end{figure}

We are just starting to see detailed measurements of the production
polarization of heavy-quark particles.  Previously,\cite{ACCMOR} data on
the heavy-quark particles which carry spin have not been sufficiently
copious to allow such determinations.  Now, E791 has measured
polarization for the open-charm $D^*$ and $\Lambda_c$ particles.  For
the $D^*$, they now have the spin-density matrix elements as functions
of $x_F$ and $p_t^2$ as shown in Fig. \ref{e791Pol}.  For the
$\Lambda_c$, the full E791 decay analysis includes the $\Lambda_c$
polarization as a function of $p_t$.\cite{E791Lpol}

Onium production polarization is also measured, and is useful in
determining the extent of color-octet contributions to the production
cross section.  Models of color-octet production predict large
polarization.  Yet, E866/NuSea observes little polarization for the
onium ground states, and quite large polarizations for the excited onium
states.\cite{E866Ypol}


The simplest models of heavy-quark production predict that the
heavy-quark particle and antiparticle will appear back-to-back in the 
center of mass.  This correlation caries over directly to the laboratory 
angle in the transverse plane between the particle and antiparticle.  It
is the simplest correlation to measure, requiring less than complete
reconstruction of the particle and antiparticle.

\begin{figure}[t]
\vspace{16.0cm}
\includegraphics{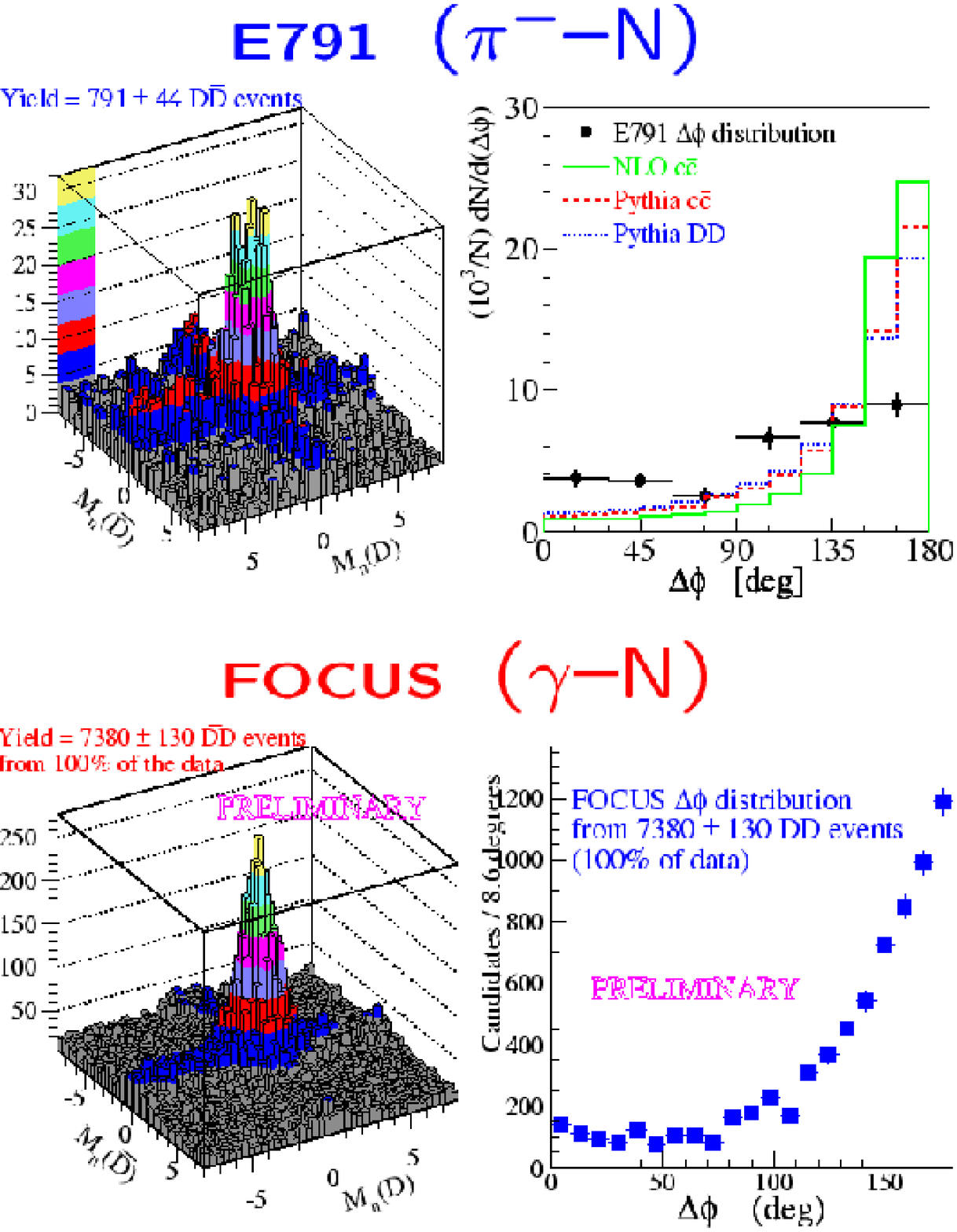}
\caption{\it
      Comparison of charm photoproduction (E831/FOCUS) and
hadroproduction (E791) particle-antiparticle correlations in the plane
transverse to the incident beam.    
\label{e831-e791corr} }
\end{figure}

The earliest observations of particle/antiparticle correlations were
made with the complete reconstruction of one charm particle, and
incomplete reconstruction of the mate.  Now, E791 and FOCUS have made
high statistics measurements with the complete reconstruction of both
particles (Fig. \ref{e831-e791corr}), reducing the uncertainties
associated with acceptance corrections on the second particle.  A broad
range of correlations has been published by E791.\cite{E791pairs}

Given time constraints, I will only discuss one correlation.  This
one leads to the distribution in the angle between the charm particle
and antiparticle in plane transverse to the beam.  The E831/FOCUS
photoproduction data (like earlier, lower statistics measurements) is
much more peaked at 180 degrees (the back-to-back angle) than the E791
hadroproduction data  (Fig. \ref{e831-e791corr}).  Thus, the hadronic
environment plays a critical role in smearing the simpler QCD
predictions of perturbative calculations and parton-shower simulations.

\section{Intrinsic $k_t$}

All the transverse momentum distributions in LO calculations tend to be
unphysical delta-functions, like the back-to-back production
discussed above.  NLO and parton shower models provide smearing due to
gluon emission, etc.  However, this smearing is not enough to match
the measured distributions like those for $p_t$ and the transverse angle
between the particle and antiparticle.\cite{E791pairs}  Several
simulation packages insert an additional effect, that due to
transverse momentum of initial partons in hadrons. This transverse
momentum is called ``intrinsic $k_t$.''

When the model intrinsic $k_t$ is varied until the simulation
predictions match data, we find that $k_t$ in the range of 1 to 3
GeV/$c$ is needed.  How can the intrinsic $k_t$ of partons inside a
proton or neutron be typically much more than the nucleon rest mass?  
Intrinsic $k_t$ must be a misnomer for something else.  But, what?

\section{Concluding Remarks}
In this review, we have seen a large amount of quality, fixed-target,
heavy-quark production data.  There is still more to come: from FOCUS,
SELEX, and COMPASS.  Each of the present experiments is providing
a healthy variety of observations.  

The major features of our understanding of this heavy-quark production
data are (1) factorization of the process, (2) the perturbative-QCD
production of the heavy quarks - dominated by gluons in the incident
hadrons, and (3) a more-or-less complicated hadronization.  The
picture works fairly well in describing the data. Nevertheless, there
are also, still, several outstanding issues.  I would mention
particularly:\\

Understanding the K-factor for charm and beauty production, including\\
\indent \indent any role of the color-octet process.

Understanding the onium cross sections and the role, if any, of the\\
\indent \indent color-octet process.

Untangling intrinsic $k_t$, and seeing it as shorthand for ... what?

Being better able to understand the relation between A-dependence \\
\indent \indent effects in fixed-target experiments and the
quark-gluon\\
\indent \indent  plasma signatures sought in heavy-ion collisions.\\

We can benefit from detailed, systematic comparisons across the range
of observations made, particularly comparing open and hidden heavy
flavor mesons, comparing mesons and baryons, and comparing the varied
spin states and particle types (i.e., with varied light valence quarks
in the final heavy-quark particle).  In addition, we must have parameter
sets which explain the full range of observations, not just parameters
which are tuned measurement-by-measurement.

Achieving a more complete understanding of the production of heavy
quarks, even at fixed-target energies, will help us understand QCD
itself, as well as help to provide guidance in our studies of the
signals and backgrounds for the even heavier objects to come at the
highest energies. 

%
%
%
\section{Acknowledgements}
I would like to give special thanks to
 Chuck Brown (NuSea),
 Dave Christian (E690),
 Peter Cooper (SELEX),
 Brad Cox (E771),
 Erik Gottschalk (E690 and FOCUS),
 David Langs (E791), Jim Russ (SELEX), and
 Panagiotis Spentzouris (E815)
for help obtaining the latest results.
I also want to thank the organizers of the very useful and enjoyable
HQ2K meeting.

\section{Comments and Questions From Attendees}
\noindent Comment from Brad Cox:\\
\indent ``Let me emphasize one point you made.  Those people who are
doing heavy ion collisions should take into account the complex A
dependences seen in many fixed-target experiments.  Only then will the
observations that they make be on solid ground.''\\
\\
\noindent Question from Ikaros Bigi:\\
\indent ``You showed very intriguing data from E791 on the polarization
of $\Lambda_c$ produced by pions.  Can one conjecture then that FOCUS
and SELEX will have sizable samples of polarized $\Lambda_c$ and
$\Xi_c$?\\
\\
\noindent Reply from Appel:\\
\indent ``Yes, I hope that we will see results from such samples.  Of
course, the polarization may be quite different with incident photon and
hyperon beams.  That's part of the interest these results should have.''


\begin{thebibliography}{99}
\bibitem{E789charmA}
E789 Collaboration, M.J. Leitch  {\it et al.},
Phys. Rev. Lett. {\bf 72}, 2542 (1994);
\bibitem{E769charmA}  E769 Collaboration, G.A. Alves {\it et al.},
Phys. Rev. Lett. {\bf 70}, 722 (1993).
\bibitem{WA82charmA}
WA82 Collaboration, M. Adamovich {\it et al.}, 
Phys. Lett. B {\bf 284} 453 (1992).
\bibitem{JAdep}
E866/NuSea Collaboration, R.E. Tribble {\it et al.}, 
Nucl. Phys. A {\bf 663} 761 (2000); 
E789 Collaboration, M.J. Leitch {\it et al.},
Phys. Rev. D {\bf 52} 4251 (1995); 
\bibitem{YAdep}
E772 Collaboration, D.M. Alde {\it et al.},
Phys. Rev. Lett. {\bf 66} 2285 (1991).
\bibitem{E769xF}     E769 Collaboration, 
G.A. Alves {\it et al.}, Phys. Rev. Lett. {\bf 77}, 2392 (1996).
\bibitem{brodsky}
R.~Vogt and S.J.~Brodsky, Nucl. Phys. B {\bf 478}, 311 (1996); 
S.J.~Brodsky {\it et al.}, Phys. Lett. B {\bf 93}, 451 (1980).
\bibitem{E769asy}  E769 Collaboration,
G.A. Alves {\it et al.}, Phys. Rev. Lett. {\bf 72}, 812 (1994);
Erratum-ibid. {\bf 72}, 1946 (1994).
\bibitem{E791asy}    E791 Collaboration,
E.M.~Aitala {\it et al.}, Phys. Lett. B {\bf 411}, 230 (1997).
\bibitem{E791tuneMC} E791 Collaboration,  
E.M.~Aitala {\it et al.}, Phys. Lett. B {\bf 371}, 157 (1996).
\bibitem{E789Jpsi}   E789 Collaboration
M.S.~Kowitt {\it et al.}, Phys. Rev. Lett. {\bf 72}, 1318 (1994).
\bibitem{E653}       E653 Collaboration
K. Kodama  {\it et al.}, Phys. Lett. B {\bf 316}, 188 (1993).
\bibitem{Berezhnoy}
A.V.~Berezhnoy, V.V.~Kiselev, and A.K.~Likhoded, hep-ph/9905555.
\bibitem{E791prod}   E791 Collaboration,
E.M.~Aitala {\it et al.}, Phys. Lett. B {\bf 462}, 225 (1999).
\bibitem{E771mumu}   E771 Collaboration,
T.~Alexopoulos {\it et al.}, Phys. Rev. Lett. {\bf 82}, 41 (2000).
\bibitem{SLDalpha}   SLD Collaboration, 
K.~Abe {\it et al.}, Phys. Rev. D {\bf 59}, 012002 (1999).  
\bibitem{E789Jpsisigma}  E789 Collaboration
M.H.~Schub {\it et al.}, Phys. Rev. D {\bf 52}, 1307 (1995).
\bibitem{E771Jpsisigma}  E771 Collaboration,
T.~Alexopoulos {\it et al.}, Phys. Rev. D {\bf 62}, 032006 (2000).
\bibitem{CDFpol}     CDF Collaboration,
T. Affolder {\it et al.}, Phys. Rev. Lett. {\bf 85}, 2886 (2000).
\bibitem{E866Adep}   E866 Collaboration,     
M.J.~Leitch {\it et al.}, Phys. Rev. Lett. {\bf 84}, 3256 (2000).
\bibitem{E781asy}    E781/SELEX Collaboration,
M. Iori {\it et al.}, 
hep-ex/9910039.
\bibitem{ACCMOR}     
M.~Jezabek, K.~Rybicki, and R.~Rylko, Phys. Lett. B {\bf 286}, 175
(1992).
\bibitem{E791Lpol}   E791 Collaboration,
E.M.~Aitala {\it et al.}, Phys. Lett. B {\bf 471}, 449 (2000).
\bibitem{E866Ypol}   E866/NuSea Collaboration,
C.N. Brown {\it et al.}, hep-ex/0011030.
\bibitem{E791pairs}  E791 Collaboration, 
E.M.~Aitala {\it et al.}, EPJdirect C {\bf 4}, 1 (1999).  
\end{thebibliography}
\end{document}